\pdfoutput=1
\documentclass[a4paper]{jacow}
\usepackage[colorlinks=true]{hyperref}
\usepackage{pdfpages,multirow,ragged2e} 
\usepackage[utf8]{inputenc}

\title{Multi-objective optimization of the dynamic aperture for the Swiss Light Source upgrade}
\author{M. Kranj\v{c}evi\'{c}\thanks{marija.kranjcevic@psi.ch}, B. Riemann, A. Adelmann, A. Streun\\
Paul Scherrer Institut, 5232 Villigen PSI, Switzerland}

\date{}

\usepackage{amsmath,siunitx,graphicx}
\usepackage{algorithm}
\usepackage[noend]{algpseudocode}
\usepackage{xcolor}
\usepackage{dcolumn}

\renewcommand{\vec}{\boldsymbol}

\begin{document}

\maketitle

\begin{abstract}
The upgrade of the Swiss Light Source, called SLS~2.0, is scheduled for 2023--24. The current storage ring will be replaced by one based on multi-bend achromats, allowing for about 30 times higher brightness. Due to the stronger focusing and the required chromatic compensation, finding a reasonably large dynamic aperture (DA) for injection, as well as an energy acceptance for a sufficient beam lifetime, is challenging. In order to maximize the DA and prolong the beam lifetime, we combine the well-known tracking code \texttt{tracy} with a massively parallel implementation of a multi-objective genetic algorithm (MOGA), and further extend this with constraint-handling methods. We then optimize the magnet configuration for two lattices: the lattice that will be used in the commissioning phase (phase-1), and the lattice that will be used afterwards, in the completion phase (phase-2). Finally, we show and further analyze the chosen magnet configurations and in the case of the phase-1 lattice compare it to a pre-existing, manually optimized solution.
\end{abstract}\vspace{0.5em}

\section{Introduction}
A new generation of ring-based synchrotron light sources based on multi-bend achromats is emerging, with many facilities being constructed or upgraded around the world. These machines have a stronger non-linear behavior of particle motion, resulting in a significantly smaller DA than their predecessors. Maximizing the DA is thus more difficult and more important. This can be done indirectly, by reducing the dominant resonance driving terms (e.g.,~\cite{johan-sls2}), or directly, by increasing the aperture and energy acceptance (e.g.,~\cite{ehrlichman,li-cheng,wang-2019}).

We apply the latter approach and use a MOGA, combined with a modified version of the well-known tracking code \texttt{tracy}~\cite{tracy}, to search for the magnet strengths that (1) maximize three transverse DAs (at different energies), as well as (2) optimize the tune footprint (and by proxy, the energy acceptance and beam lifetime). We will explain this in more detail in the following two sections, and then show the results for two lattice configurations of the SLS upgrade~\cite{andreas-ipac20}.\vspace{0.5em}

\section{Optimization problem}

\subsection{Dynamic Aperture (DA)}
For an energy $\delta$, we compute the transverse DA, denoted $\mathrm{DA}_\delta$, using $2K$ evaluation rays that are equally distributed in angle. Along the evaluation line corresponding to the angle $\theta_k = k\pi/K$, $k\in\{0,\dots,2K-1\}$, the linear aperture, i.e., the aperture of the lattice with all nonlinear elements zeroed, has some length $\bar L_k$. Denoting by $L_{k,\delta}$ the length of the DA along this line, we define the \emph{line objective} $f_{k,\delta}$ as
$  f_{k,\delta} = \left( \max\{0,\bar L_k - L_{k,\delta}\} / \bar L_k \right)^2.$
Note that the DAs larger than the linear aperture are not rewarded.
Assuming that the magnetic lattice structure is planar,
\begin{align*}
  \mathrm{DA}_\delta &= \frac{1}{2K} \sum_{k=0}^{2K-1} f_{k,\delta} = \frac{1}{2K} \left( f_{0,\delta} + f_{K,\delta} + 2 \sum_{k=1}^{K-1} f_{k,\delta}\right).
\end{align*} 
Note that $\mathrm{DA}_\delta \in [0,1]$.
This is a slight modification of the approach presented in~\cite{ehrlichman}.
We use the biased binary search as shown in Algorithm~\ref{alg:biased-binary}, with $\varepsilon = 10^{-5}$, to compute $L_{k,\delta}$. A preliminary computation showed that $b = 0.25$ is a good choice, reducing the run time relative to $b=0.5$ by $25\%$.
\begin{algorithm}
\caption{Biased binary search}\label{alg:biased-binary}
\begin{algorithmic}[1]
\State $r_+ \leftarrow \bar L_k$, $r_- \leftarrow 0$ \label{alg:bb-initialize}
\While {$r_+ - r_- \geq \varepsilon$} \label{alg:bb-stopping-criterion}
\State $r \leftarrow b \cdot r_- + (1-b)\cdot r_+$ \label{alg:bb-evaluate}
\If {$r$ is stable}
$r_- \leftarrow r$
\Else 
{ $r_+ \leftarrow r$}
\EndIf
\EndWhile
\State $L_{k,\delta} \leftarrow \left(r_- + r_+\right)/2$
\end{algorithmic}
\end{algorithm}

\subsection{Chromatic Tune Footprint}
We sample the tune footprint $\vec q$ in $P = 51$ equidistant values of the energy $\delta$ in $[-\delta_\text{max},\delta_\text{max}]$ with $\delta_\text{max} = 0.05$, and denote these values by $\vec q_i=\vec q(\delta_i)$.
In the particle tracking code \texttt{tracy} it can happen that the computation of $\vec q$ breaks down for some values of $\delta$, so we define
\[ \texttt{sqd} = \sum_{\stackrel{i = 1}{\vec q_i\text{ computable}}}^P g(\vec q_i),\]
where $g(\vec q)$ is the squared Euclidean distance of $\vec q$ to 
the triangle formed by three intersecting 2nd order resonances around the on-momentum tune (see Table~\ref{Tab:dynap_b038_run0}, first row, gray lines). E.g., for the phase-1 lattice this triangle is $(39,15)-(39.5,15.5)-(39.5,15)$. Note that $g(\vec q) = 0$ when $\vec q$ is inside of the triangle.
Furthermore, similarly to~\cite{ehrlichman}, we define
$\texttt{unstable}_\pm = 1 - |\delta_{u,\pm}|/\delta_\text{max},$
where $\delta_{u,+}$ and $\delta_{u,-}$ denote the first (i.e., smallest in magnitude) positive and negative values, respectively, that are located outside the triangle or not computable.\\

To sum up, the optimization problem we solve is
\begin{align*}
& \underset{\boldsymbol{d} = (d_1,\dots,d_N)}{\text{min}}
&& \big(\underbrace{\mathrm{DA}_{-0.03}}_{F_1},\underbrace{\mathrm{DA}_{0}}_{F_2},\underbrace{\mathrm{DA}_{0.03}}_{F_3},\underbrace{\texttt{unstable}_{\mp}}_{F_4,F_5}\big),\\
& \text{subject to} && \texttt{sqd} = 0. 
\end{align*}

\subsection{Search Space}

The \emph{design point} $\boldsymbol d  = (d_1,\dots,d_N)$ contains sextupole and possibly also octupole strengths. The physical limits for the sextupole and integrated octupole strengths are $\SI{650}{m^{-3}}$ and $\SI{350}{m^{-3}}$, respectively. To have the horizontal and vertical chromaticity $\xi_x,\xi_y \in [0,1]$, we consider these to be design variables and use them to tune two of the sextupole strengths, denoted $\vec t = (t_1,t_2)$, using $\vec \xi = \mathbf M \vec m + \mathbf T \vec t + \vec \xi_\text{ua}$. $\vec \xi_\text{ua}$ is the chromaticity of the unaltered lattice, $\vec m$ are the remaining sextupole strengths, $\vec \xi = (\xi_x,\xi_y)$, and $\mathbf T$ and $\mathbf M$ are (constant) tuning matrices (all known).
When only sextupoles are used, $N=9$. When also octupoles are used, $N=17$, so we reduce this number to $N=13$ by treating the dual magnet families that fulfill similar functionalities as one magnet family.

\section{Optimization method}

\subsection{Multi-Objective Genetic Algorithm (MOGA)}
A design point $\boldsymbol{d}_1$ \emph{dominates} $\boldsymbol{d}_2$ if it is not worse in any of the objectives, and it is strictly better in at least one objective. We use a massively parallel implementation of a MOGA~\cite{IneichenPhD,Neveu2019,PhysRevAccelBeams.22.122001} to find points that are not dominated by any other point, called \emph{Pareto optimal points}. The basic steps are in Algorithm~\ref{alg:MOGA}. 
The design points, called \emph{individuals} in the context of a MOGA, comprising the first generation are chosen uniformly at random from the given intervals (line~\ref{alg:MOGA-initialize}).
They are then evaluated, i.e., their objective function values are computed using \texttt{tracy} (line~\ref{alg:MOGA-evaluate}). Afterwards, a number of cycles is performed, each resulting in a new generation (lines~\ref{alg:MOGA-cycle}--\ref{alg:MOGA-selector2}). In every cycle, new individuals are created using two operators: crossover and mutation (lines~\ref{alg:MOGA-crossover1}--\ref{alg:MOGA-crossover2}), and then evaluated (line~\ref{alg:MOGA-evaluate-new}). Approximately $M$ fittest individuals are chosen to comprise the new generation (line~\ref{alg:MOGA-selector2}). This process is parallelized such that a new generation is created (line~\ref{alg:MOGA-selector2}) once $n$ new individuals have been evaluated (line~\ref{alg:MOGA-evaluate-new}).
\begin{algorithm}[!b]
\caption{Multi-objective genetic algorithm}\label{alg:MOGA}
\begin{algorithmic}[1]
\State random population of individuals, $\boldsymbol{I}_i$, $i = 1, \dots, M$ \label{alg:MOGA-initialize}
\State evaluate the population \label{alg:MOGA-evaluate}
\While {a stopping criterion not reached} \label{alg:MOGA-cycle}
\For {pairs of individuals $\boldsymbol{I}_i$, $\boldsymbol{I}_{i+1}$} \label{alg:MOGA-crossover1}
\State crossover($\boldsymbol{I}_i$, $\boldsymbol{I}_{i+1}$), mutate($\boldsymbol{I}_i$), mutate($\boldsymbol{I}_{i+1}$) \label{alg:MOGA-crossover2}
\EndFor
\State evaluate new individuals \label{alg:MOGA-evaluate-new}
\State choose $M$ fittest individuals for the next generation \label{alg:MOGA-selector2}
\EndWhile
\end{algorithmic}
\end{algorithm}

\subsection{Constraint Handling}

\subsubsection{Implicit ranking}
Almost all randomly chosen individuals that comprise the first generation are \emph{infeasible}, i.e., their tuning sextupoles are outside of the bounds. In this case we avoid computing the objective function values (the run time for the tuning procedure is negligible)
and set ($m = \SI{650}{m^{-3}}$)
$F_i \leftarrow 2 + \max\left\{0,|t_i| - m\right\}$ for  $i=1,2$
and 
$F_i \leftarrow 2$ for $i = 3,4,5$, 
i.e., infeasible individuals are compared based on the severity of their constraint violations.

\subsubsection{Penalty function}
To enforce the constraint $\texttt{sqd} = 0$, for every feasible individual we set $F_i \leftarrow F_i + \texttt{sqd}/2$. 
If $\texttt{sqd} \geq 2$, the tune footprint extends so far outside the triangle that it cannot be considered better than all infeasible points. Therefore, we allow the possibility that the penalized objectives of this feasible individual are compared with constraint violations of an infeasible individual. This either results in the feasible individual being chosen (standard behavior) or the infeasible individual being chosen (in which case its tuning sextupoles are likely close to the admissible bounds).

\section{Results}

To find a good magnet configuration for the phase-1 lattice we ran an optimization with $M=300$, $n=100$. The 34th generation already contained only feasible individuals. On 108 processes of Intel~Xeon~E5\_2680v3 this took almost \SI{12}{min}, while reaching the 1000th generation took \SI{15}{h}~\SI{38}{min}. 
We show the objective function values for the manually optimized solution, which we call the \emph{design solution}, as well as for a good individual found in generation 2968 in Table~\ref{Tab:objectives}, and give additional information in Table~\ref{Tab:dynap_b038_run0}.
For the phase-2 lattice we used $M=3000$ and $n = 1000$. 
A point found in generation 972 is shown in Tables~\ref{Tab:objectives} and \ref{Tab:dynap_b038_run0}. 

\begin{table}[h!]
\small
\centering
\caption{Objective function values computed with 500 turns in \texttt{tracy} for a few magnet configurations with $\texttt{sqd} = 0$.}
\label{Tab:objectives}
\begin{tabular}{lccccc}
\toprule
\textbf{Objective} & $\boldsymbol{F_1}$ & $\boldsymbol{F_2}$ & $\boldsymbol{F_3}$ & $\boldsymbol{F_4}$ & $\boldsymbol{F_5}$ \\ \midrule
phase-1 design~sol.  & $0.03$ & $0.004$ & $0.010$ & $0$ & $0$ \\
phase-1 optim.~sol.  & $0.02$ & $0.001$ & $0.005$ & $0$ & $0$ \\ \midrule
phase-2 optim.~sol.  & $0.5$ & $0.03$ & $0.2$ & $0.16$ & $0$ \\ \bottomrule
\end{tabular}
\end{table}

\section{Conclusions}
We used a multi-objective genetic algorithm to find a good dynamic aperture and energy acceptance for the Swiss Light Source upgrade. The optimization method and its implementation can easily be applied to similar problems. Furthermore, the method and code can be enhanced in the following ways.
First, the convergence of the method could be further improved, e.g., using surrogate models. Second, a more accurate model could be used, including, e.g., quadrupole knobs and cavities. Third, the sensitivity of the solution with respect to misalignments could be included.

\section{Acknowledgements}
M.~Aiba provided the manually optimized configuration for the phase-1 lattice. We executed the computations on the Euler cluster~\cite{euler} of ETH~Zurich at the expense of a PSI~grant.

\begin{table*}
    \centering
    \caption{The first row shows the chromatic tune footprint, and the remaining three the DA at $-0.03$, $0$ and $0.03$, recomputed in OPA~\cite{opa}, for the phase-1 design solution (left), and the optimized solutions for the phase-1 (center) and phase-2 (right) lattices. In the case of the phase-1 lattice, both points have a full energy acceptance ($-0.05$ to $0.05$). The increased DA areas for the optimized solution relative to the design solution are consistent with the relationship of the corresponding objective function values in Table 1. It should be noted that the DAs of the design solution were optimized beyond the linear aperture limits. In case of the phase-2 lattice, a design solution with a sufficient energy acceptance does not exist. The energy acceptance of the optimized solution shown in the right column is $-0.04$ to $0.05$.}
    \label{Tab:dynap_b038_run0}
    \setlength{\tabcolsep}{2pt} 
    \begin{tabular}{r|ccc}
& phase-1 design solution & phase-1 optimized solution & phase-2 optimized solution \\ \hline & & & \\[-.8em]
\rotatebox{90}{\qquad\quad chromatic tune footprint}&
\includegraphics[width=.63\columnwidth]{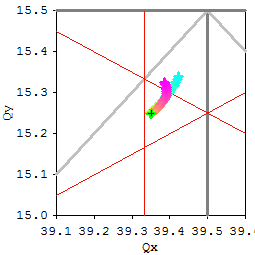}&
\includegraphics[width=.63\columnwidth]{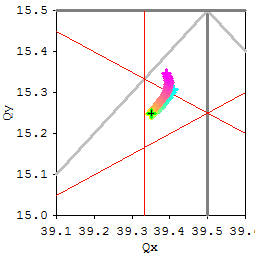}&
\includegraphics[width=.63\columnwidth]{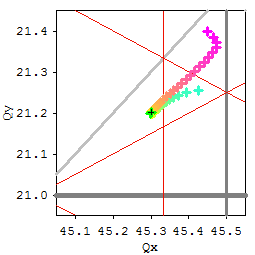}\\
\rotatebox{90}{\qquad\qquad\, DA, $\delta=-0.03$} &
\includegraphics[width=.63\columnwidth]{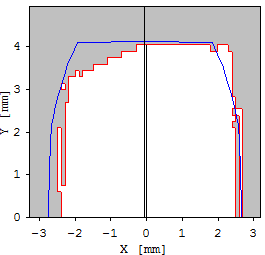}&
\includegraphics[width=.63\columnwidth]{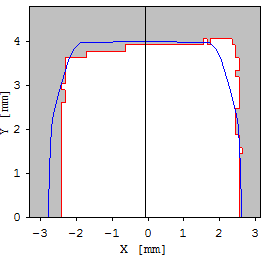}&
\includegraphics[width=.63\columnwidth]{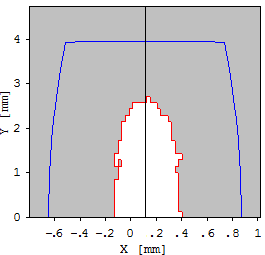}\\
\rotatebox{90}{\qquad\qquad\quad DA, $\delta=0$} &
\includegraphics[width=.63\columnwidth]{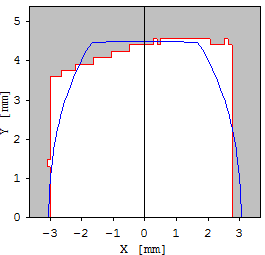}&
\includegraphics[width=.63\columnwidth]{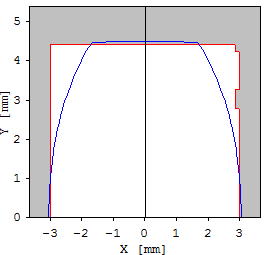}&
\includegraphics[width=.63\columnwidth]{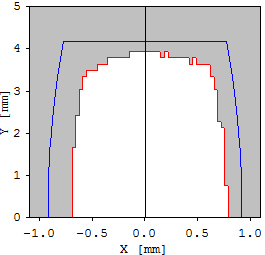}\\
\rotatebox{90}{\qquad\qquad\, DA, $\delta=0.03$} &
\includegraphics[width=.63\columnwidth]{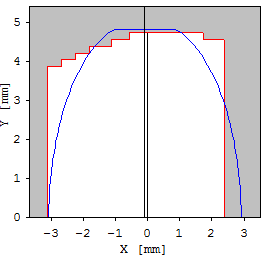}&
\includegraphics[width=.63\columnwidth]{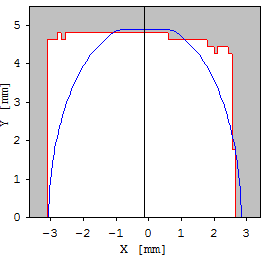}&
\includegraphics[width=.63\columnwidth]{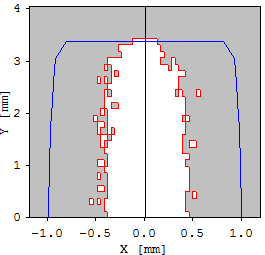}
\end{tabular}
\end{table*}


\clearpage

\begin{thebibliography}{10}

\bibitem{johan-sls2}
J.~Bengtsson and A.~Streun.
\newblock \href{https://ados.web.psi.ch/SLS2/Notes/SLS2-BJ84-001.pdf}{Robust design strategy for {SLS 2}}.
\newblock Report SLS2-BJ84-001, PSI, 2017.

\bibitem{ehrlichman}
M.~P. Ehrlichman.
\newblock \href{https://doi.org/10.1103/PhysRevAccelBeams.19.044001}{Genetic algorithm for chromaticity correction in diffraction limited storage rings}.
\newblock {\em Phys. Rev. Accel. Beams}, 19:044001, 2016.

\bibitem{li-cheng}
Y.~Li et~al.
\newblock \href{https://doi.org/10.1103/PhysRevAccelBeams.21.054601}{Genetic algorithm enhanced by machine learning in dynamic aperture optimization}.
\newblock {\em Phys. Rev. Accel. Beams}, 21:054601, 2018.

\bibitem{wang-2019}
F.~Wang et~al.
\newblock {Machine learning for design optimization of storage ring nonlinear dynamics}.
\newblock arXiv:1910.14220, 2019.

\bibitem{tracy}
Tracy-3, {S}elf-consistent charged particle beam tracking code based on a
  symplectic integrator by {J}.~{B}engtsson, available for download at
  \url{https://github.com/jbengtsson/tracy-3.5}.

\bibitem{andreas-ipac20}
A.~Streun et~al.
\newblock The lattice for the upgrade of the {S}wiss {L}ight {S}ource {SLS}~2.0.
\newblock accepted for {\em IPAC'20}, Caen, France, 2020.

\bibitem{IneichenPhD}
Y.~Ineichen.
\newblock \href{https://doi.org/10.3929/ethz-a-009792359}{Toward massively parallel multi-objective optimization with application to particle accelerators}.
\newblock {\em ETH Research Collection}, (Diss.\ 21114), 2013.

\bibitem{Neveu2019}
N. Neveu et~al.
\newblock \href{https://doi.org/10.1103/PhysRevAccelBeams.22.054602}{Parallel general purpose multiobjective optimization framework with application to electron beam dynamics}.
\newblock {\em Phys. Rev. Accel. Beams}, 22:054602, 2019.

\bibitem{PhysRevAccelBeams.22.122001}
M.~Kranj\v{c}evi\'{c} et~al.
\newblock \href{https://doi.org/10.1103/PhysRevAccelBeams.22.122001}{Constrained multiobjective shape optimization of superconducting rf cavities considering robustness against geometric perturbations}.
\newblock {\em Phys. Rev. Accel. Beams}, 22:122001, 2019.

\bibitem{euler}
\url{https://scicomp.ethz.ch/wiki/Euler}.

\bibitem{opa}
{OPA} accelerator optics software.
\newblock \url{https://ados.web.psi.ch/opa}.

\end{thebibliography}
\end{document}